\documentclass[11pt,twoside]{article}

\newcommand{\EQ}{\begin{equation}}
\newcommand{\EN}{\end{equation}}
\newcommand{\EQA}{\begin{eqnarray}}
\newcommand{\ENA}{\end{eqnarray}}

\newcommand{\Fig}[1]{Fig.~\ref{#1}}
\newcommand{\FFig}[1]{Figure~\ref{#1}}

\newcommand{\bra}[1]{\langle #1\rangle}

{}
{}
{}

{}
{}
{}
{}
{}
{}
{}
{}
{}
\newcommand{\meanBB}{\overline{\mbox{\boldmath $B$}}{}}{}
{}
{}
{}
{}
{}
{}
{}
{}
{}

{}

{}

{}
{}

\newcommand{\uu}{\mbox{\boldmath $u$} {}}

\newcommand{\bb}{\mbox{\boldmath $b$} {}}
\newcommand{\BB}{\mbox{\boldmath $B$} {}}

\def\Pm{\mbox{\rm Pr}_M}
\def\Rm{\mbox{\rm Re}_M}

\def\Rey{\mbox{\rm Re}}

\def\kf{k_{\rm f}}

\def\urms{u_{\rm rms}}

\def\Beq{B_{\rm eq}}

\newcommand{\kG}{\,{\rm kG}}

\newcommand{\Mm}{\,{\rm Mm}}

\newcommand{\yapj}[3]{ #1, {ApJ,} {#2}, #3}

\newcommand{\yapjs}[3]{ #1, {ApJS,} {#2}, #3}
\newcommand{\yan}[3]{ #1, {Astron.\ Nachr.,} {#2}, #3}

\newcommand{\yana}[3]{ #1, {A\&A,} {#2}, #3}

\newcommand{\yjfm}[3]{ #1, {J.\ Fluid Mech.,} {#2}, #3}

\newcommand{\yphy}[3]{ #1, {Physica,} {#2}, #3}
\newcommand{\yaraa}[3]{ #1, {ARA\&A,} {#2}, #3}

\newcommand{\yprl}[3]{ #1, {Phys.\ Rev.\ Lett.,} {#2}, #3}

\newcommand{\ynat}[3]{ #1, {Nature,} {#2}, #3}

\newcommand{\ysph}[3]{ #1, {Solar Phys.,} {#2}, #3}

\newcommand{\ypre}[3]{ #1, {Phys.\ Rev.\ E,} {#2}, #3}

\newcommand{\yjour}[4]{ #1, {#2}, {#3}, #4}

\newcommand{\yproc}[5]{ #1, in {#3}, ed.\ #4 (#5), #2}

\newcommand{\ppapj}[3]{ #1, {ApJ}, {#2}, to be published in the #3 issue}

\usepackage{asp2006}
\usepackage{epsf}
\usepackage{psfig}
\usepackage{lscape}
\usepackage{graphicx}

\begin{document}
\title{From fibril to diffuse fields during dynamo saturation}
\author{Axel Brandenburg}
\affil{Nordita, Roslagstullsbacken 23, 10691 Stockholm, Sweden
}

\begin{abstract}
The degree of intermittency of the magnetic field of a large-scale dynamo
is considered.
Based on simulations it is argued that there is a tendency for the field
to become more diffuse and non-intermittent as the dynamo saturates.
The simulations are idealized in that the turbulence is strongly helical
and shear is strong, so the tendency for the field to become more diffuse
is somewhat exaggerated.
Earlier results concerning the effects of magnetic buoyancy are discussed.
It is emphasized that the resulting magnetic buoyancy is weak compared with
the stronger effects of simultaneous downward pumping.
These findings are used to support the notion that the solar dynamo
might operate in a distributed fashion where the near-surface shear layer
could play an important role.
\end{abstract}

\section{Introduction}

In the early days of dynamo theory the degree of intermittency of the
generated magnetic field was not much of an issue.
However, with the development of mean-field theory it became clear that
the magnetic field can be thought of as consisting of a mean component
together with a fluctuating one.
The fluctuating component was initially thought to be weak, but that too
changed when it was realized that at large magnetic Reynolds numbers
the fluctuations can strongly exceed the level of the mean field.

Given the intermittent nature of solar magnetograms, the surface magnetic
field can well be described as fibril.
This description was introduced by Parker (1982) to emphasize that such
a field may have rather different properties than a more diffuse field.
The fibril nature of the magnetic field is particularly well illustrated
by the fact that sunspots are relatively isolated features covering
only a small fraction of the solar surface.
It is often assumed that the fibril magnetic field structure extends also
into deeper layers.
On the other hand, observations of sunspots suggest that spots are rather
shallow phenomena (Kosovichev 2002).
Furthermore, simulations of turbulent dynamos tend to show that
the dynamo-generated magnetic field becomes less fibril as the
fraction of the mean to the total magnetic field increases.
Such dynamos are generally referred to as large-scale dynamos
(as opposed to small-scale dynamos) and they require either kinetic
helicity or otherwise some kind of anisotropy.
These ingredients are generally assumed to be present in the Sun,
and they are also vital for many types of mean-field dynamos,
in particular the $\alpha\Omega$-type dynamos.
It is therefore of interest to study in more detail the dependence of
the degree of intermittency of the field on model parameters.

The significance of looking at the degree of intermittency of the Sun's
magnetic field is connected with the question of how important is magnetic
buoyancy in transporting mean magnetic field upward to the surface and
out of the Sun (Moreno-Insertis 1983).
Magnetic buoyancy may therefore act as a possible saturation mechanism
of the dynamo (see, e.g., Noyes et al.\ 1984), with the consequence of
nearly completely wiping out magnetic fields of equipartition strength
within the convection zone.
On the other hand, if magnetic buoyancy is not a dominant effect, the dynamo
may operate in a much more distributed fashion (Brandenburg 2005).

\section{Fully helical dynamos}

\begin{figure}[t!]
\centering\includegraphics[width=\textwidth]{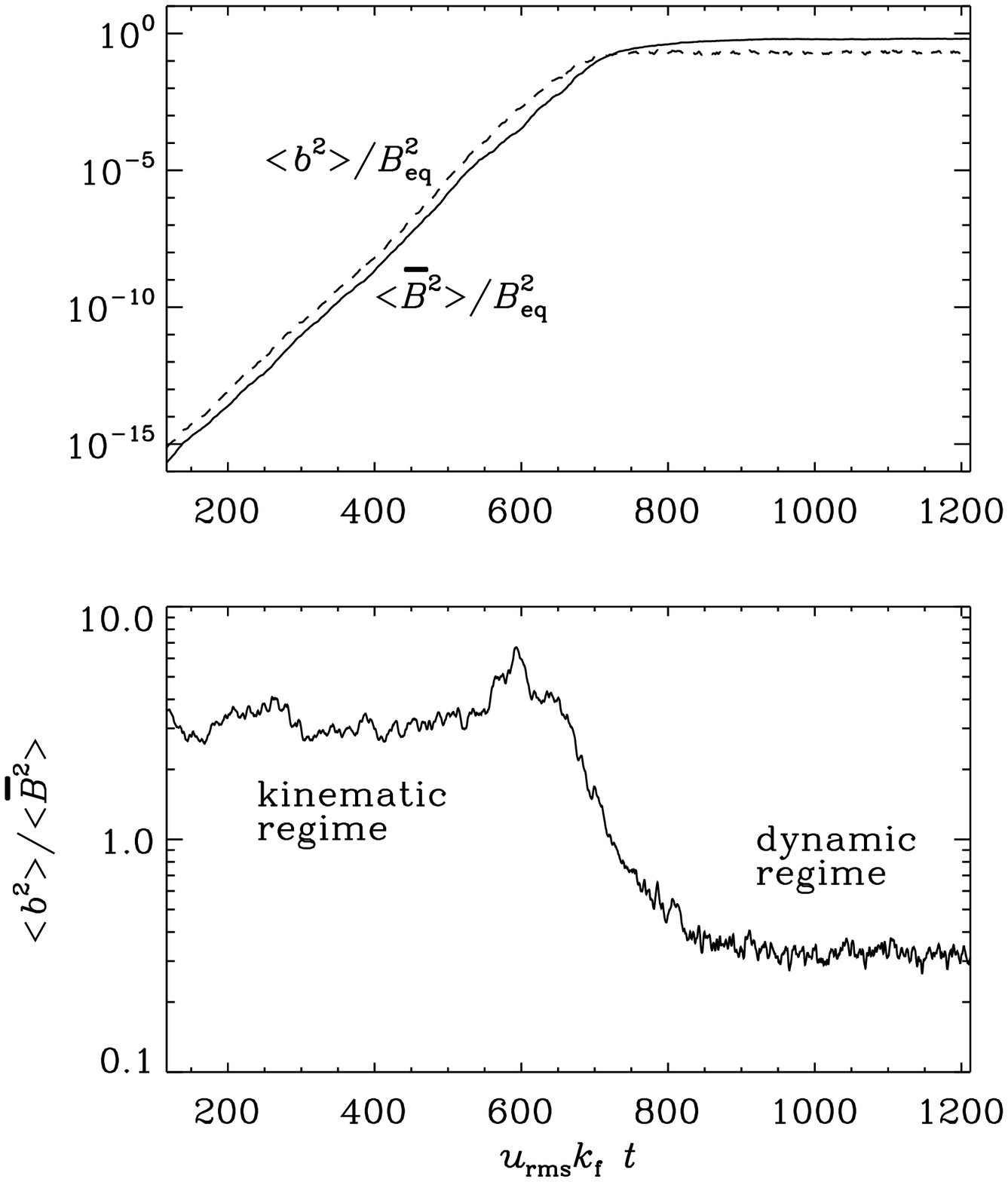}\caption{
Dependence of the normalized mean-squared value of the mean field
$\bra{\meanBB^2}$ (solid line in the upper panel) and the fluctuating field
$\bra{\bb^2}$ (dashed line in the upper panel) and their ratio
(lower panel) for $\Rm=6$ and $\Pm=0.04$.
The equipartition field strength $\Beq=\bra{\mu_0\rho\uu^2}$ has been
introduced for normalization purposes.
}\label{pbmean_fluct}\end{figure}

Let us begin by looking at an idealized case of a dynamo in
the presence of fully helical forcing.
We shall distinguish between the kinematic regime where the field is
weak and still growing exponentially, and the dynamic regime where the
field is strong and beginning to reach saturation field strength.
In \Fig{pbmean_fluct} we plot the dependence of the mean-squared
values of the small-scale and large-scale fields defined here by
horizontal averages, so $\BB=\meanBB+\bb$, where $\meanBB$ and
$\bb$ have been defined as the mean and fluctuating fields.
Note that in the kinematic regime the energy of the magnetic fluctuations
exceeds that of the mean field by a factor of about 3, while in the
dynamic regime this ratio is only about 1/3.
Here we have used data from a recent paper  of Brandenburg (2009) were
the magnetic Reynolds number is only about 6, while the fluid Reynolds
number is 150, so the magnetic Prandtl number is 0.04.
The turbulence is forced with a maximally helical forcing function
at a wavenumber of about 4 times the minimal wavenumber of the domain.
This ratio is also called the scale separation ratio and it also determines
the ratio of magnetic fluctuations to the mean field in the kinematic regime,
and its inverse in the dynamic regime (Blackman \& Brandenburg 2002).

\begin{figure}[t!]
\centering\includegraphics[width=\textwidth]{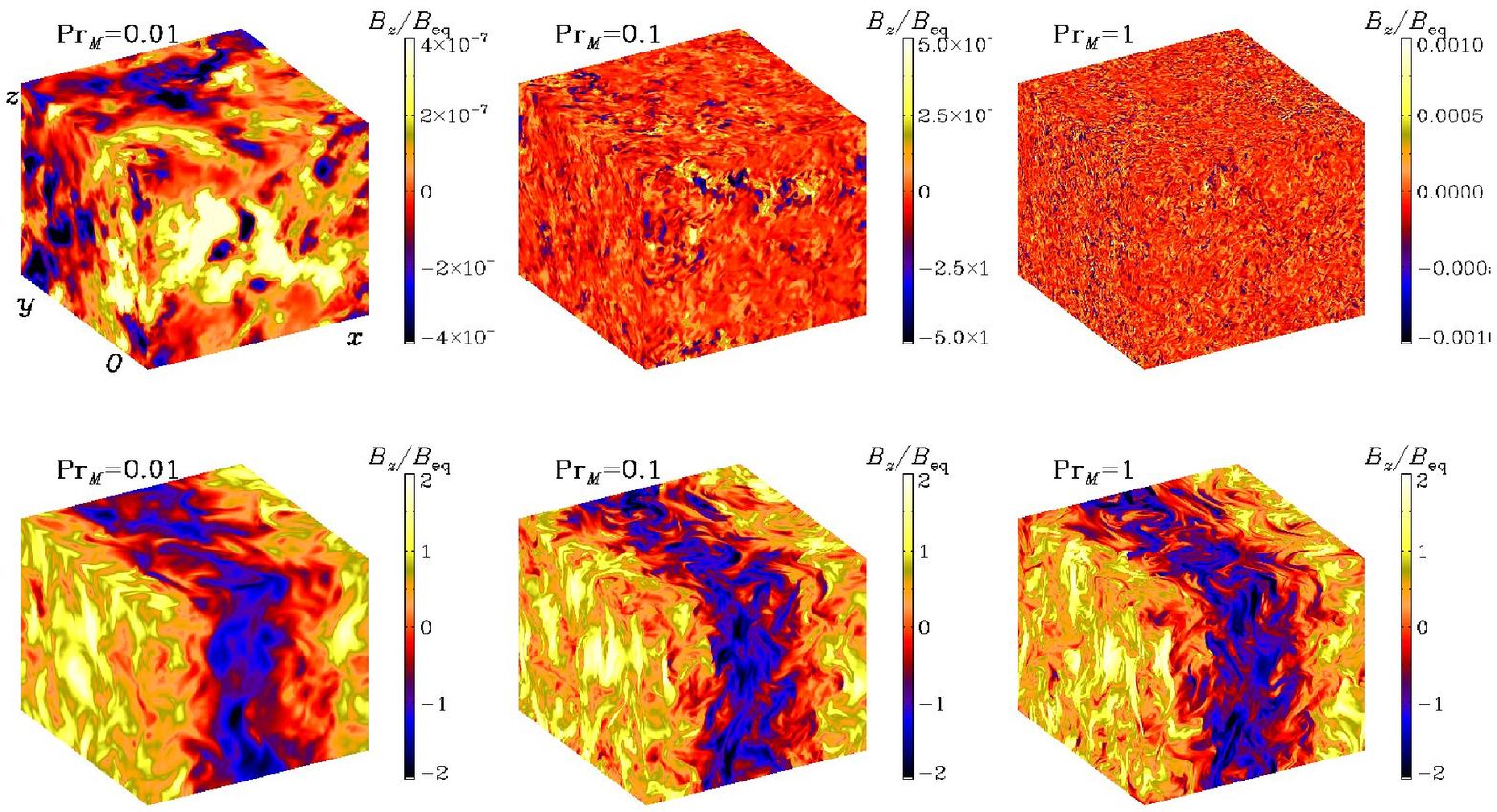}\caption{
Visualization of one component of the magnetic field in the kinematic
regime (upper row) compared with the saturated regime (lower row)
for magnetic Prandtl numbers ranging from $\Pm=0.01$ to 1 at $\Rey=670$.
The orientation of the axes is indicated for the first panel,
and is the same for all other panels.
Adapted from Brandenburg (2009).
}\label{BB1}\end{figure}

Depending on the value of the magnetic Prandtl number $\Pm$,
i.e.\ the ratio of kinematic viscosity to magnetic diffusivity,
the field can be rather intermittent and lack large-scale order,
especially when the magnetic Prandtl number is not small; see \Fig{BB1}.
Note however the emergence of a large-scale pattern in the kinematic stage
for $\Pm=0.01$, while for $\Pm=0.1$ there are only a few extended patches
and for $\Pm=1$ the field is completely random and of small scale only.
However, when the dynamo saturates, a large-scale structure
emerges regardless of the value of the magnetic Prandtl number
and the field is considerably less intermittent than in the early
kinematic stages.
These simulations (Brandenburg 2009) were used to argue that in the
Sun, where $\Pm$ is very small, the onset of large-scale dynamo action
should not depend on the actual value of $\Pm$, even though the onset
of small-scale dynamo action does depend on it (Schekochihin et al.\
2005; Iskakov et al.\ 2007).

\begin{figure}[t!]
\centering\includegraphics[width=\textwidth]{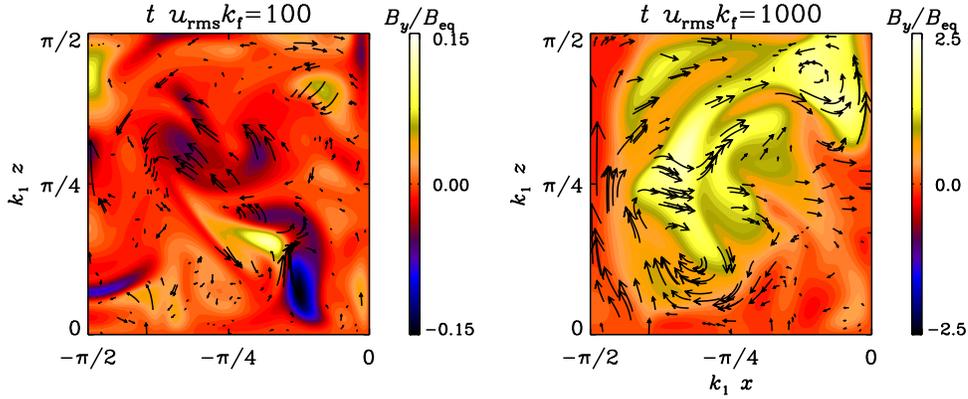}\caption{
Snapshots of the magnetic field in the meridional plane
during the kinematic stage ($t=100$ turnover times, left panel) and the
saturated stage ($t=1000$ turnover times, right panel).
Vectors in the meridional plane are superimposed on a color/gray scale
representation of the azimuthal field.
The color/gray scale is symmetric about red/mid-gray shades, so the absence
of blue/dark shades (right panel) indicates the absence of negative values.
Note the development of larger scale structures during the saturated
stage with basically unidirectional toroidal field.
Adapted from Brandenburg (2005).
}\label{pslice_128b3}\end{figure}

Another example is forced turbulence in the presence of a systematic
shear flow that resembles that in low latitudes of the solar
convection zone and open boundary conditions at the surface and
the equator.
Such a model was studied by Brandenburg \& Sandin (2004)
to determine how the $\alpha$ effect is modified in the presence of
magnetic helicity fluxes, and by Brandenburg (2005) in order to
determine the structure of dynamo-generated magnetic fields.
In \Fig{pslice_128b3} we compare meridional cross-sections of the
toroidal component of the magnetic field at a kinematic time
($t\urms\kf=100$) with that at a later time when the dynamo has
saturated and a large-scale field has developed ($t\urms\kf=1000$),
where $\urms$ is the turbulent rms velocity and $\kf$ is the wavenumber
of the energy-carrying eddies or the forcing wavenumber in this case.

The case shown in \Fig{pslice_128b3} looks like the magnetic Reynolds
number is small, but this is not really the case.
In fact, the magnetic Reynolds number based on the inverse wavenumber
of the energy-carrying eddies, $\urms/\eta\kf$, is about 80.
Here, $\kf/k_1=5$ is the forcing wavenumber in units of the smallest
wavenumber in the domain, $k_1=2\pi/L$, where $L$ is the toroidal extent
of the computational domain.
So, the magnetic Reynolds number based on $L$, which is sometimes also quoted,
would then be about $2\pi\times5\approx30$ times larger, i.e.\ about 2400.

Note also that, unlike the early kinematic stage when there can still
be many sign reversals, at later times the field points mostly in the
same direction.
Indeed, the toroidally averaged magnetic field captures about 50\% to
70\% of the total magnetic energy in the saturated state.

These simulations confirm that there is a clear tendency for the magnetic
field to become less intermittent and more space-filling and diffuse as
the dynamo saturates.
It must be noted, however, that these simulations are idealized in that the
turbulence is driven by a forcing function that is maximally helical, and
that the shear is relatively strong, i.e.\ the shear-flow amplitude is about
five times stronger than the rms velocity of the turbulence.
In the Sun this ratio is about unity.
Therefore one must expect that the degree to what extent the field tends
to become more diffuse is in reality less strong than what is indicated
by the simulations presented here.

\section{Magnetic buoyancy}

\begin{figure}[t!]\centering
\includegraphics[width=0.99\textwidth]{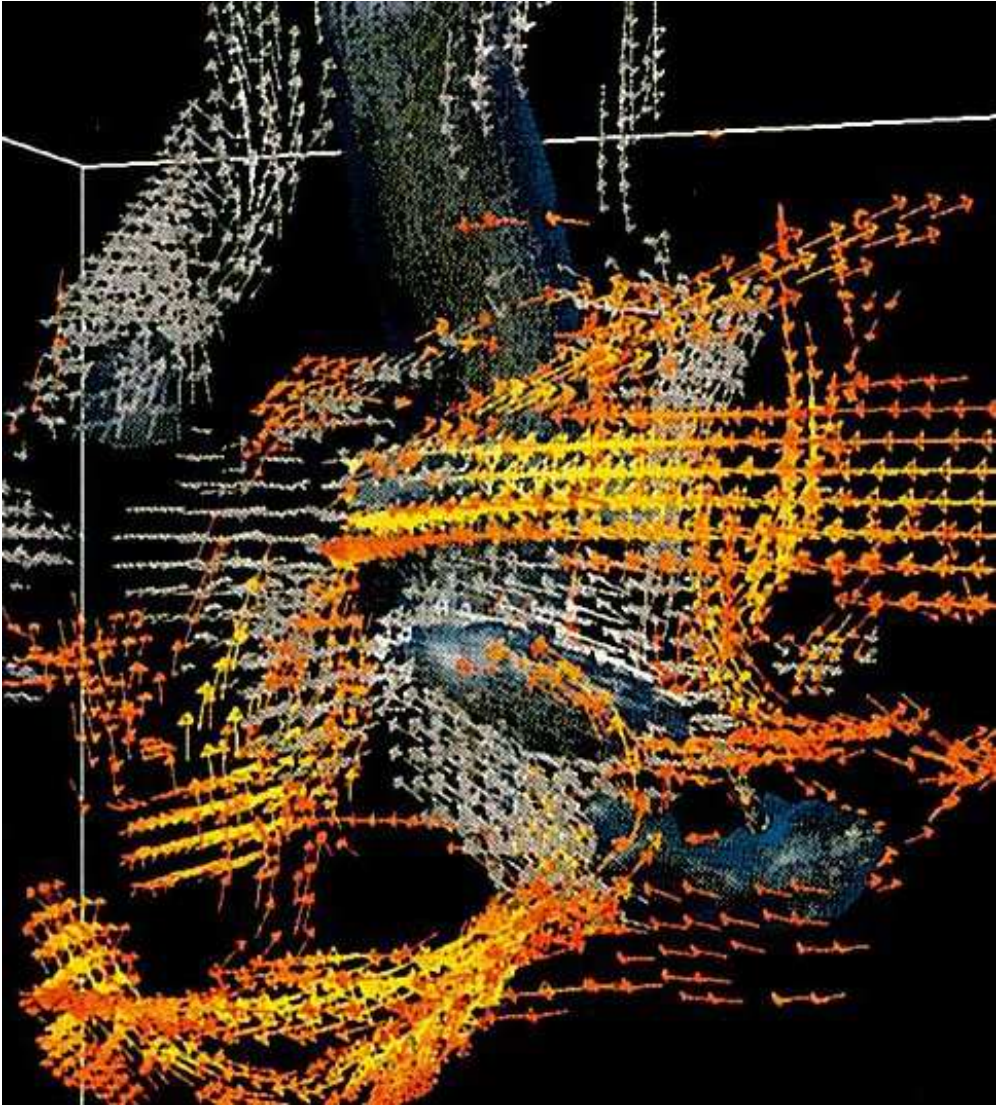}\caption{
Snapshot from a video animation showing magnetic field vectors in yellow
(the strongest) and orange (less strong) together with those of vorticity
in white.
Transparent surfaces of constant negative pressure fluctuations are shown
in blue.
Note that the vectors of magnetic field form flux tubes that get wound
up around a tornado-like vortex in the middle.
Adapted from Brandenburg \& Tuominen (1991).
}\label{BT91}\end{figure}

In the early 1980s, dynamo theory was confronted with the issue of magnetic
buoyancy (Spiegel \& Weiss 1980).
It was thought that buoyant flux losses would reduce the dynamo efficiency.
This effect was then also built into dynamo models of various types
as a possible saturation mechanism (Noyes et al.\ 1984, Jones et al.\ 1985,
Moss et al.\ 1990).
However, with the first compressible simulations of turbulent dynamo action
(Nordlund et al.\ 1992) it became clear that magnetic buoyancy is subdominant
compared with the much stronger effect of turbulent downward pumping.
\FFig{BT91} shows a snapshot from a video animation of magnetic field vectors
together with those of vorticity (Brandenburg \& Tuominen 1991).
The magnetic field forms flux tubes that get wound
up around a tornado-like vortex in the middle.

\begin{figure}[t!]\centering
\includegraphics[width=0.99\textwidth]{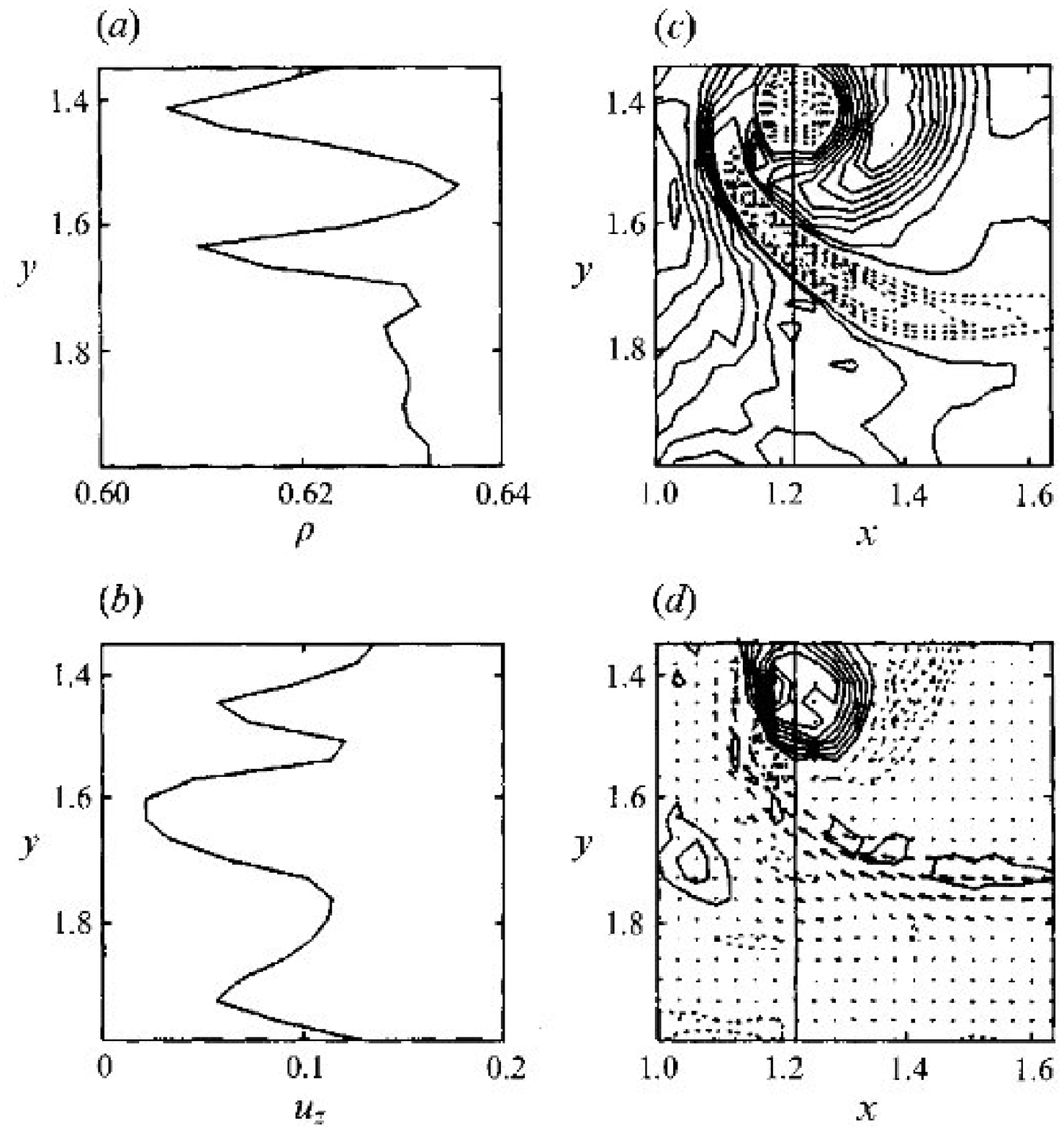}\caption{
Panel (a) shows a horizontal density profile through the line indicated
in panel (c) by the line in the corresponding horizontal cross-section
through a strong magnetic flux tube.
Dotted contours indicate values less than the average.
Panel (b) gives the vertical velocity (positive values mean upward
motion) along the line indicated in the horizontal cross-section
in panel (d).
Here the vectors indicate magnetic field vectors.
Adapted from Brandenburg et al.\ (1996).
}\label{BJNRRT96}\end{figure}

In \Fig{BJNRRT96} this magnetic buoyancy of the flux tubes is analyzed
in more detail.
This figure confirms that there is indeed magnetic buoyancy, but it is
balanced in part by the effects of downward pumping and the explicit
downward motion in the proximity of the downdraft where the field
is most strongly amplified during its descent.

\begin{figure}[t!]\begin{center}
\includegraphics[width=.49\textwidth]{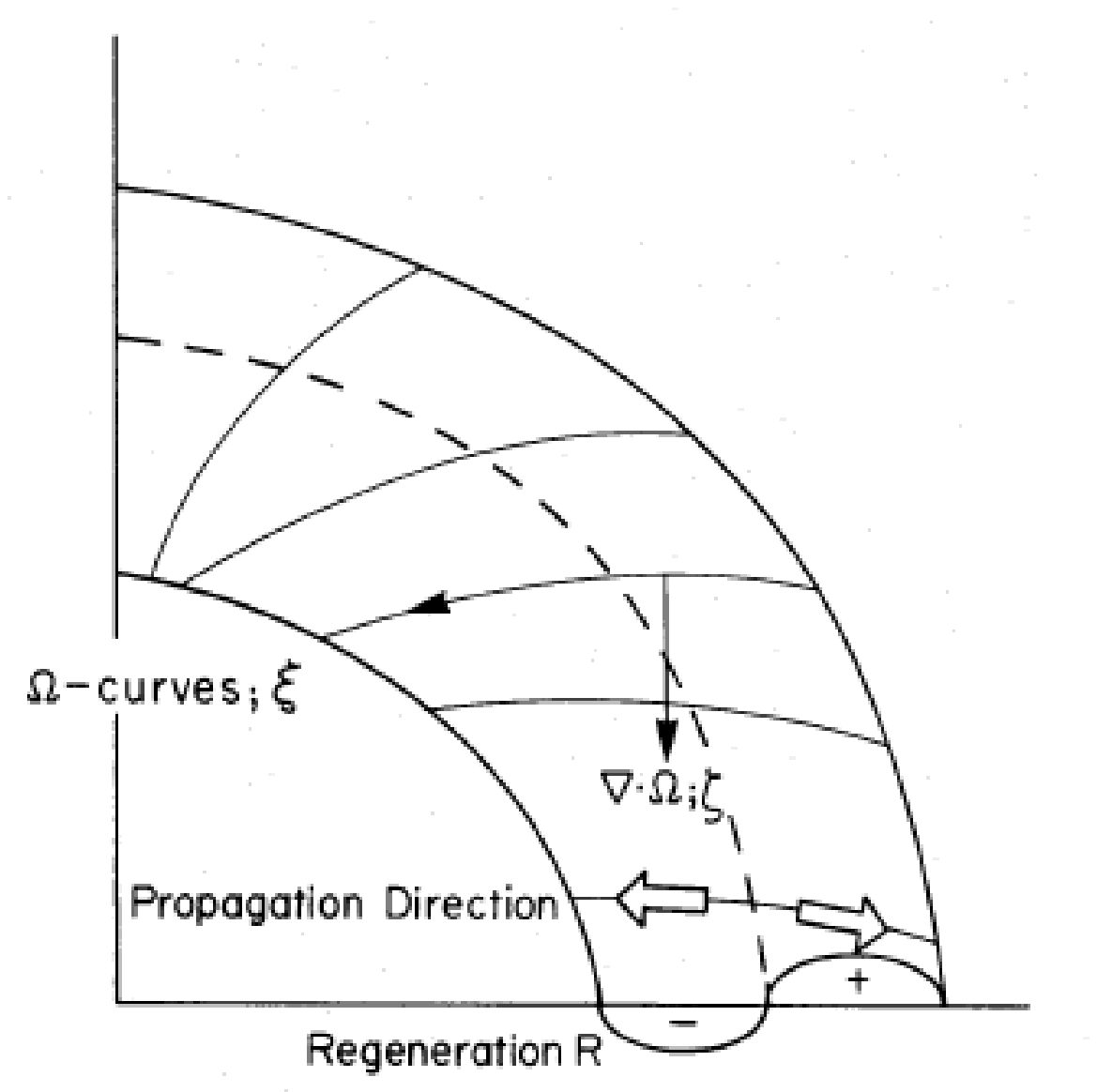}
\includegraphics[width=.49\textwidth]{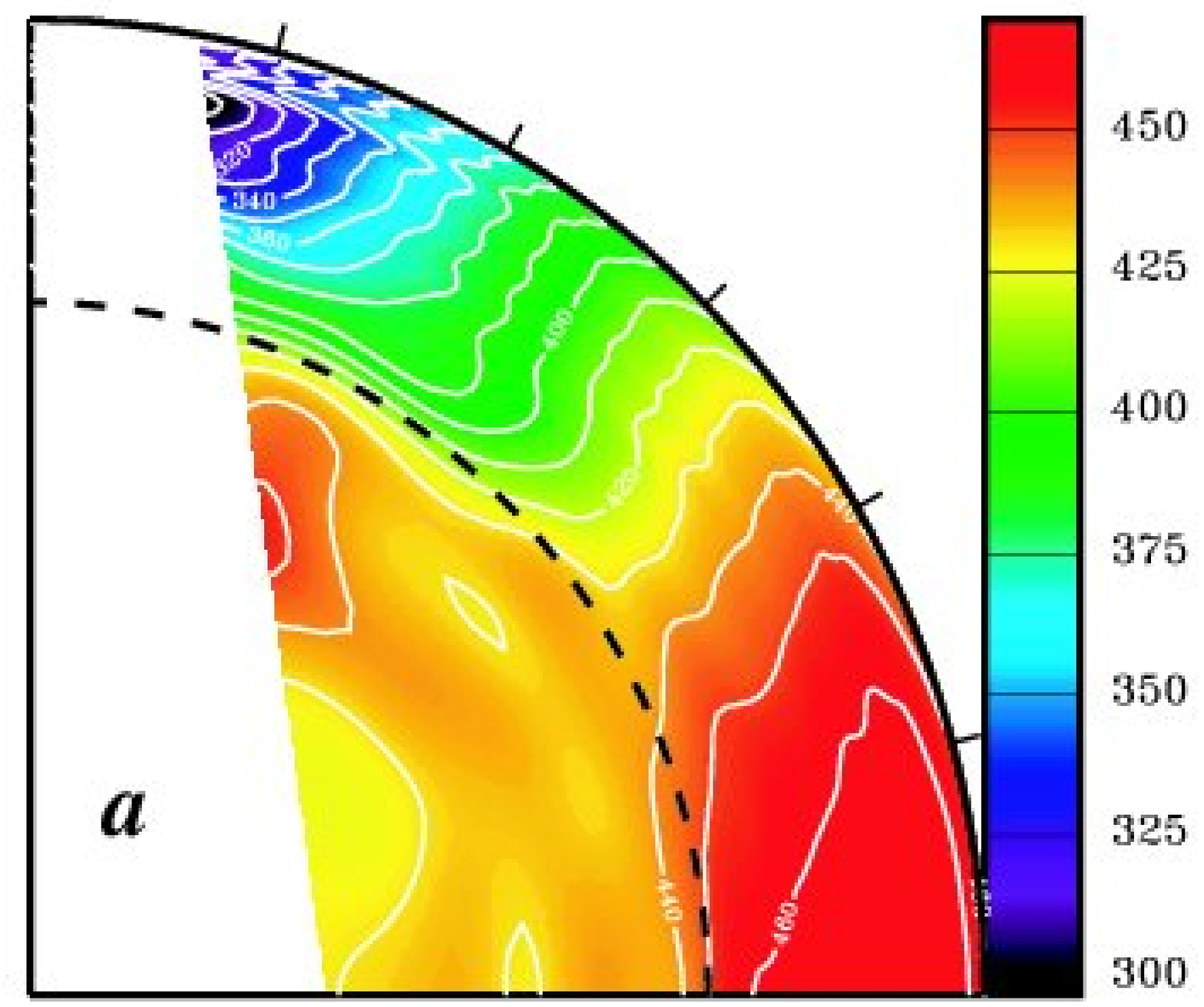}
\end{center}\caption[]{
Comparison of the differential rotation contours that were originally
expected by Yoshimura (1975) based on solar dynamo model considerations
(left) with those by Thompson et al.\ (2003) using helioseismology (right).
Note the similarities between the contours on the left (over the bulk
of the convection zone) and those on the right (over the outer 5\% of
the solar radius).
}\label{pdiffrot}\end{figure}

\begin{figure}[t!]\begin{center}
\includegraphics[width=.99\textwidth]{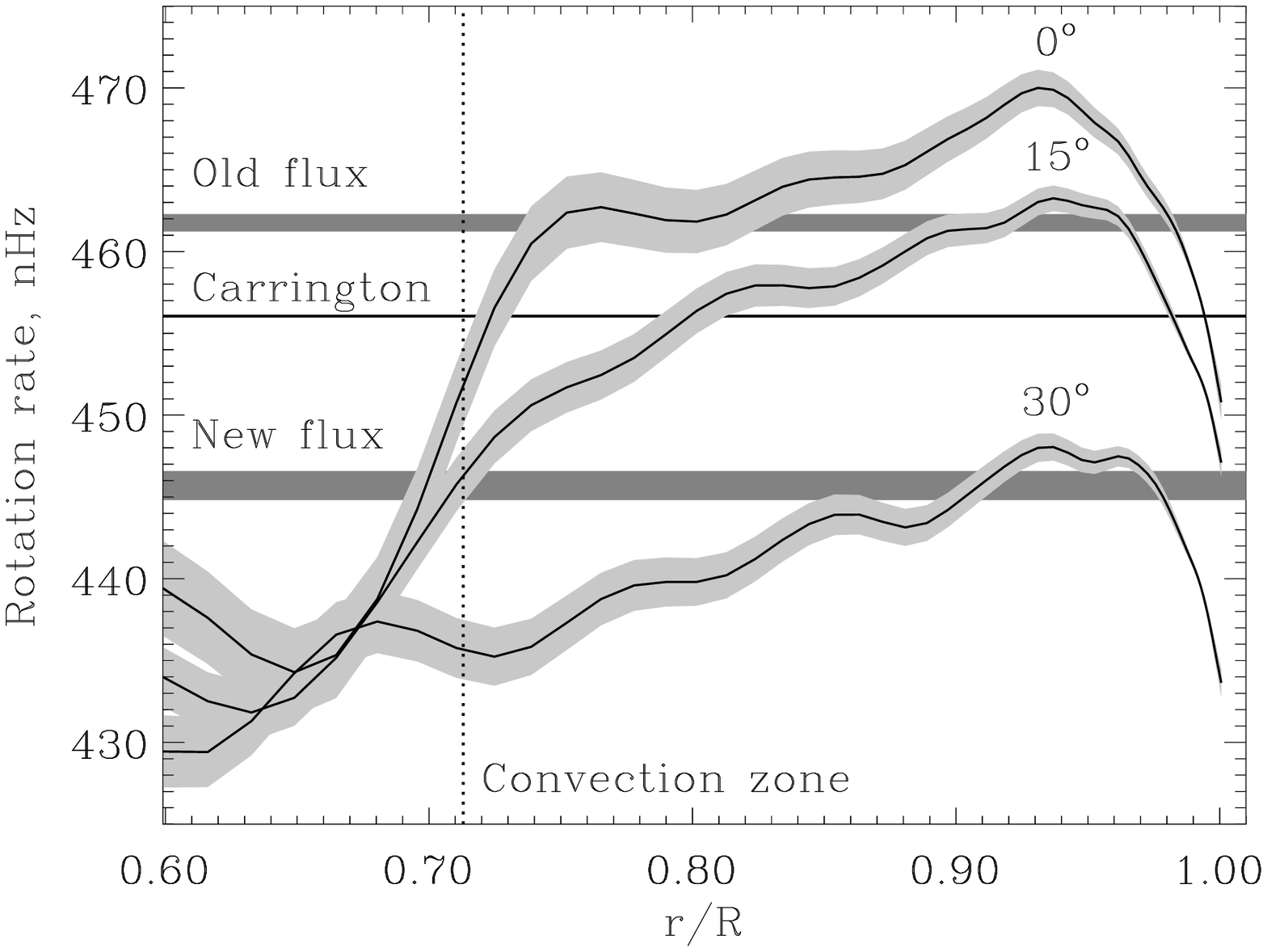}
\end{center}\caption[]{
Radial profiles of angular velocity
as given by Benevolenskaya et al.\ (1999).
Note the sharp negative radial gradient near the surface.
}\label{bene_et98}\end{figure}

\section{Connection with distributed dynamos}

We have discussed the nature of the magnetic field of a large-scale
dynamo in the saturated regime and have argued that the field becomes
diffuse and more nearly space-filling as the dynamo saturates and that
the effects of magnetic buoyancy are weak compared with the downward
motions associated with the strong downdrafts in convection.
Here we have mostly focused on earlier simulations, but it is
important to realize that at the moment there is no general
agreement about the detailed nature of the solar dynamo.
Is it essentially of $\alpha\Omega$ type, or are there other more
dominant effects responsible for generating a large-scale magnetic field?
What causes the equatorward migration of the toroidal magnetic flux belts?
Is it the dynamo wave associated with the $\alpha\Omega$ dynamo, or
is it the meridional circulation that overturns the intrinsic
migration direction (Choudhuri et al.\ 1995; Dikpati \& Charbonneau 1995).
What is the dominant shear-layer in the Sun for the $\alpha\Omega$ dynamo
to work?
There is first of all latitudinal shear, which is the strongest in absolute
terms, and important for amplifying toroidal magnetic field as well as
promoting cyclic dynamo action (Guerrero \& de Gouveia Dal Pino 2007).
In addition, there is radial shear which might be important for determining
the migration direction of the toroidal flux belts.
However, it is not clear whether the relevant component here is
the positive $\partial\Omega/\partial r$ in the bulk or the bottom
of the convection zone, or the negative $\partial\Omega/\partial r$
at the bottom of the convection zone at higher latitudes.
Or is it the negative $\partial\Omega/\partial r$ in the near-surface
shear layer?

An attractive property of the latter proposal is that it would allow
for a dynamo scenario that is in many respects similar to that envisaged
in the early years of mean-field dynamo theory (Steenbeck \& Krause 1969,
K\"ohler 1973, Yoshimura 1975).
In \Fig{pdiffrot} we show the structure of $\Omega$ contours as they
were estimated by Yoshimura (1975) based on the constraint that the
internal angular velocity matches the latitudinal differential rotation
at the surface and that $\partial\Omega/\partial r$ is negative in the
interior so that the dynamo wave propagates equatorward.
The relative strength of the negative $\Omega$ gradient near the surface is
truly amazing and is best seen in a plot of Benevolenskaya et al.\ (1998),
which shows the radial dependence of $\Omega$ at different latitudes
(\Fig{bene_et98}).
The fact that the radial gradient is so strong is in principle not new.
Indeed, a mismatch between the higher helioseismic results for $\Omega$
some $40\Mm$ below the surface and the lower values from Doppler
measurements of the photospheric plasma was recognized since the 1980s,
but it is only now that helioseismology can actually provide detailed
data points nearly all the way to the surface.

We emphasize that these proposals ignore the possibility that the
meridional circulation could in principle turn the direction of
propagation around and might produce equatorward migration even
with a positive $\partial\Omega/\partial r$ (Choudhuri et al.\ 1995,
Dikpati \& Charbonneau 1999).
However, this requires that the induction effects given by
$\alpha$ and the radial differential rotation are separated
in space, just as it is the case for the Babcock-Leighton dynamo effect.
Although such a hypothesis was already made by Steenbeck \& Krause (1969)
for other reasons, it is not clear that this is or will be compatible
with results of turbulence simulations.

\section{Unexplored effects}

There are two important issues that need to be clarified in the context
of distributed dynamos.
One is connected with the question why the dynamo might work efficiently
in the near-surface shear layer in spite of the opposing effects of
downward pumping, for example.
The other is related to the formation of active regions and sunspots
in models lacking strong fields of $\sim100\kG$ strength at the bottom
of the convection zone, as is expected based on Joy's law and results
from the thin flux tube approximation (Chou \& Fisher 1989;
Choudhuri \& D`Silva 1990).

Regarding the first issue one might expect that it could be connected
with magnetic helicity conservation, which is now recognized as a
major culprit in causing so-called catastrophic quenching of large-scale
dynamo effects (see Brandenburg \& Subramanian 2005 for a review).
Alleviating such catastrophic quenching is facilitated by magnetic
helicity fluxes connected with scales that are shorter than those
of the large-scale field of the 11-year cycle.
Disposing of such excess magnetic helicity should be easier near the
surface than deeper down, making the near-surface shear layer more
preferred for dynamo action.
Regarding the formation of active regions and sunspots, some important
clues have been obtained by investigating mean-field turbulence effects
both in the momentum and in the energy equations.
We refer here to the work of Kitchatinov \& Mazur (2000) who find that
a self-concentration of magnetic flux is possible as a result of the
magnetic suppression of the turbulent heat flux.
Another mechanism might be connected with negative turbulent magnetic
pressure effects; see Rogachevskii \& Kleeorin (2007) for a recent
reference on this subject.
Clarifying these questions would be critical before further pursuing the
idea of distributed dynamo action in the Sun.

\acknowledgements

This work was supported in part by the Swedish Research Council,
grant 621-2007-4064, and the European Research Council under the
AstroDyn Research Project 227952.

\end{document}